# On the Transcriptomic Signature and
# General Stress State Associated with Aneuploidy

Hung-Ji Tsai[a], Anjali R. Nelliat[b], Andrei Kucharavy[c], Mohammad Ikbal Choudhury[d], Sean X. Sun[c], Michael C. Schatz[e], Rong Li[f,g,*]

[a]Institute of Microbiology and Infections,
School of Biosciences, University of Birmingham, Birmingham B15 2TT, United Kingdom
[b]Department of Systems Biology,
Harvard Medical School, Boston, MA 02115
[c]Distributed Computing Laboratory, School of Computerand communication Sciences,
Ecole Polytechnique Federale de Lausanne, Lausanne, Switzerland
[d]Department of Mechanical Engineering,
Johns Hopkins University, Baltimore, MD, 21218
[e]Department of Computer Science, Johns Hopkins University, Baltimore, MD, 21218
[f]Center for Cell Dynamics, Department of Cell Biology,
Johns Hopkins University School of Medicine, Baltimore, MD 21205
[g]Mechanobiology Institute and Department of Biological Science,
National University of Singapore, Singapore 117411

[*]To whom correspondence should be addressed:
rong@jhu.edu

Terhorst *et al.* (1) reinvestigated the common aneuploidy gene expression (CAGE), explicitly disputing the conclusion of Tsai *et al.* (2), where we used RNAseq to identify CAGE without karyotype bias for understanding the general consequence of aneuploidy. CAGE led to our initial hypothesis that aneuploid cells experience hypo-osmotic-like stress, which was further validated through extensive measurements of cell stiffness and turgor pressure, endocytosis, size, density and cytoplasmic diffusion. The observed changes were explained quantitatively with a model based on thermodynamic principles. Our conclusion was also supported by a parallel genetic screen. Terhorst *et al.* mainly contested the initial RNAseq analysis. Our major concerns are:

I) Since valid comparative transcriptomic analysis relies on carefully standardized experimental conditions, we employed isogenic aneuploid and euploid populations grown under identical conditions and processed in parallel. Surprisingly, Terhorst *et al.* compared our RNAseq data from cultures grown in a high-throughput format, used to maintain aneuploid karyotype diversity, with a dataset from the ancestral strain RLY4388 grown under a conventional condition for strain authentication. With this highly questionable comparison, authors concluded that CAGE resulted from an Environmental Stress Response (ESR) in euploid populations (3). We, but not Terhorst *et al.*, validated differentially expressed genes using qPCR on cultures grown in more conventional formats (2).

II) CAGE is independent of ESR: If we remove 108 ESR genes from CAGE gene-set, the expression of remaining 114 CAGE genes still positively correlates with hypo-osmotic stress response (Spearman's rank correlation: 0.55, *p*-value: 4.29E-8).

III) Terhorst *et al.* relied on single-sample gene set enrichment analysis (ssGSEA) as the main tool for comparing gene expression signatures. However, ssGSEA score is subject to many assumptions and requires appropriate normalization to derive gene ranking order (4). For meaningful comparison across different samples, it is more straightforward to compare expression values of parallelly processed samples as in our study.

IV) Several results in Terhorst *et al.* actually confirmed our findings: Slt2 MAP kinase activation, an indicator of cell wall stress, was higher in aneuploid cells, but Slt2-phosphorylation level was lower than that after cell wall damage - consistent with our EM data showing a lack of general cell wall defect in aneuploids (2). Interestingly, CAGE was evident in Terhost *et al*. even with the revised growth condition (5, Figure 2E), but further bootstrapping was applied to remove statistical difference (1). Additionally, the multiple dilutions/regrowth in the revised growth condition likely reduced the diversity of aneuploid populations, which we intentionally avoided.

Finally, our biophysical model predicted a nonmonotonic relationship between cell size and ploidy, consistent with the experimental observation. This was different from the linear relationship between ESR and ploidy (1). In fact, osmotic theory (6) would predict ribosome loss, if reducing cellular protein content as implied in (1), to reduce cell size but not necessarily density, in contrast to larger and lighter aneuploid cells (2). Other studies had also questioned whether ESR is an obligatory aneuploid signature (7–10). Thus, we stand by our finding that hypo-osmotic-like stress characterizes a general state of aneuploid cells.